\documentclass[10pt,twocolumn]{article}

\usepackage[margin=1in]{geometry}
\usepackage[medium,compact]{titlesec}
\usepackage{alltt}

\newif\ifpdf
\ifx\pdfoutput\undefined
  \pdffalse
  \pdfoutput=0
\else
  \ifx\pdfoutput\relax
    \pdffalse
    \pdfoutput=0
  \else
    \ifnum\pdfoutput=0
      \pdffalse
    \else
      \pdftrue
    \fi
  \fi
\fi

\ifpdf      \usepackage{hyperref}
      \usepackage[pdftex]{graphics}
      \usepackage{epstopdf}
\else\usepackage{graphics}
     \usepackage[hyphens]{url}
     \usepackage{cite}
\fi

\usepackage[T1]{fontenc}
\usepackage{times}
\usepackage{textcomp}

\usepackage[tight]{subfigure}
\usepackage{amsmath,amsthm}
\usepackage{epsfig}
\usepackage{cite}
\usepackage{color}
\usepackage{xspace}

\newtheorem{prop}{Property}

\newcommand{\note}[1]{}

\newcommand{\eat}[1]{}

\makeatletter
\let\@copyrightspace\relax
\makeatother

\usepackage{fancyhdr}

\title{
\vspace{-1.2em}
\bf\Large CloneCloud: Boosting Mobile Device Applications\\Through Cloud Clone Execution
\vspace{-0.7em}
}

\author{Byung-Gon Chun$^\dagger$, Sunghwan Ihm$^\star$, Petros Maniatis$^\dagger$, Mayur Naik$^\dagger$\\
        \textit{$^\dagger$Intel Labs Berkeley, $^\star$Princeton University}}

\date{}

\renewenvironment{thebibliography}[1]{%
    \begin{oldthebibliography}{#1}%
    \setlength{\parskip}{0ex}%
    \setlength{\itemsep}{0ex}%
}%
{%
    \end{oldthebibliography}%
}

\begin{document}

\maketitle

\begin{abstract}
Mobile applications are becoming increasingly ubiquitous and
provide ever richer functionality on mobile devices.
At the same time, such devices often enjoy strong
connectivity with more powerful machines ranging from laptops
and desktops to commercial clouds. This paper presents the design and implementation of CloneCloud, a system that automatically transforms
mobile applications to benefit from the cloud. The system is 
a flexible application partitioner and 
execution runtime that enables unmodified mobile applications running 
in an application-level virtual machine to seamlessly
off-load part of their execution from mobile devices onto device
clones operating in a computational cloud.
CloneCloud uses a combination of static analysis and dynamic
profiling to optimally and automatically partition an application so
that it migrates, executes in the cloud, and re-integrates
computation in a fine-grained manner that makes efficient use
of resources.
Our evaluation shows that CloneCloud can achieve up to 21.2x speedup of 
smartphone applications we tested and it allows different partitioning
for different inputs and networks.
\end{abstract}

\section{Introduction}

Mobile cloud computing is the next big thing. In recent research 
done by ABI research~\cite{ABI}, it has predicted that by the end of 
2014 mobile cloud computing will deliver annual revenues of 20 billion dollars.
Mobile devices as simple as phones and as complex as mobile Internet
devices with Internet access via multiple technologies, camera(s), GPS,
and other sensors are the current computing wave, competing heavily with
desktops and laptops for market and popularity.  The variety of
flash-popular applications being featured on various on-line application
stores like those of Apple, Google, Microsoft and others mean that
mobile users have no shortage of interesting things to do with their
devices, for a low fee or even free.  

This blossoming of the mobile application market is pushing mobile users
beyond the usual staples of personal information management and music
playback. Now mobile users look up songs by audio samples; play games;
capture, edit, and upload video; analyze, index, and aggregate
their mobile photo collections; analyze their finances; and manage their
personal health and wellness. Also, new rich media, mobile augmented 
reality, and data analytics applications that require heavy computation
are emerging.
Such applications recruit
increasing amounts of computation, storage, and communications from a
still limited supply on mobile devices---certainly compared to tethered, grid-powered
devices like desktops and laptops---and an extremely limited supply of
energy.  As a result, mobile applications end up in one of two camps:
1) they are either designed for the lowest common denominator device,
pushing most functionality at a service provider's site, and leaving
little computing done at the device as a thin client; or 2) they are built monolithically to
run on the device, taking a long time to execute on low-end devices,
even when a split client-server design might have been desired.

Fortunately, such devices often enjoy  strong connectivity, especially in
developed areas. What is more, there is increasingly broad availability of tethered computing,
storage, and communications to spare on commercial clouds, at nearby wireless hotspots
equipped with computational resources (e.g., cloudlet~\cite{cloudlet}), or
at the user's PC and plugged-in laptop. Putting these two
trends together, we recently made the case
for a flexible architecture that enables the seamless use of
\emph{ambient} computation to augment mobile device
applications~\cite{Chun2009}.  In this paper, we take a first step
towards realizing this vision, by designing and implementing the first
version of the \emph{CloneCloud} system.

\begin{figure}
\centerline{\includegraphics{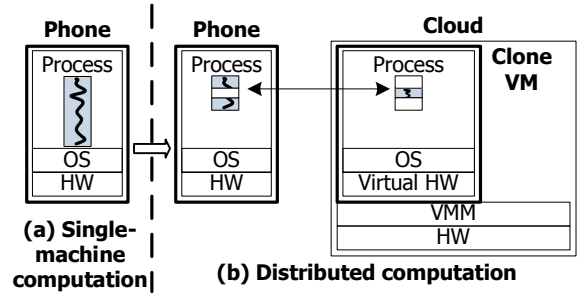}}
\caption{\small CloneCloud system model. CloneCloud 
transforms a single-machine execution (mobile device
computation) into a distributed execution (mobile device and cloud computation) 
automatically.}
\label{fig:approach}
\end{figure}

CloneCloud boosts unmodified mobile applications by seamlessly
off-loading part of their execution from the mobile device onto
\emph{device clones} operating in a computational cloud\footnote{Throughout this paper, we use the term "cloud" in a broader sense to include diverse ambient computational resources discussed above.}.
It is designed to serve as a platform for generic mobile-device processing
as a service.  Conceptually, our system automatically transforms a
single-machine execution (e.g., computation on a smartphone) into a
distributed execution that is optimal given the network connection to
the cloud, if needed, the relative processing capabilities of the mobile device 
and cloud, and the application's computing patterns
(Figure~\ref{fig:approach}).

The underlying motivation for such a system lies in the following
intuition: as long as execution on the cloud is significantly faster
than execution on the mobile device (or more reliable, more secure,
etc.), paying the cost for sending the relevant data and code from the
device to the cloud and back may be worth it.  Unlike partitioning a
service by design between an undemanding mobile client and a
computationally expensive server in a provider's infrastructure,
CloneCloud late-binds this kind of partitioning. 
Only when the metric (e.g., performance or energy) of the newly partitioned 
application is better than that of the existing application, it makes
sense to partition an application.
In practice, the partitioning decision
may be more fine-grained than a yes/no answer (i.e., it may result in
carving off different \emph{amounts} of the original application for
cloud execution). Furthermore, the decision may be impacted not only by
the application itself, but also by the expected workload and the
execution conditions, such as network connectivity and CPU speeds of
both mobile and cloud devices.
A fundamental design goal for CloneCloud is to allow such
fine-grained flexibility on what to run where, which traditional
client-server partitionings hard-wire early on in the development
process.

Another design goal for CloneCloud is to take the 
programmer out of the business of application partitioning.  While we
conjecture that automatic partitioning is unlikely to produce optimized
applications that can rival what a competent programmer would hand-code,
we assert that competent programmers are also unlikely to willingly do
such a hand-coding job for every possible set of circumstances a user
may face.  The
kinds of applications on mobile platforms that are featured on
application stores and gain flash popularity tend to be low-margin
products, whose developers have little incentive to optimize manually
for different combinations of architectures, network conditions, battery
lives, and hosting infrastructures.  
Consequently, CloneCloud aims to make application partitioning seamless, 
and based only on the deployed version of the application, without need for
source code.

Our work in this paper applies primarily to
application-layer virtual machines, such as the Java VM, DalvikVM from
the Android Platform, and
Microsoft's~.NET. The relative ease of manipulating
application executables and migrating pieces thereof to computing
devices of diverging architectures made the AppVM model a
promising first platform on which to explore our work.
We expect some---but not all---of our design decisions to
carry over when addressing such partitioning at lower layers in the
execution stack, e.g., to UNIX-level processes, to kernel-level process
containers, or to mobile hypervisors.

\begin{figure}
\centering
\includegraphics[width=\columnwidth]{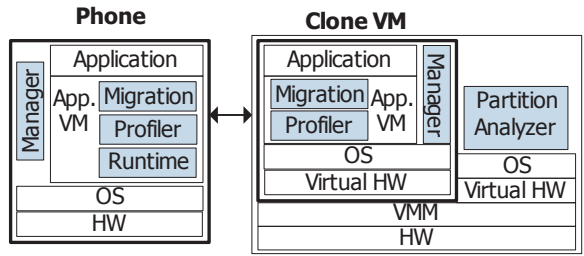}
\caption{The CloneCloud prototype architecture.}
\label{fig:architecture}
\end{figure}

The CloneCloud prototype described here meets all our design
goals, by rewriting an unmodified application executable. While the
modified executable runs, at automatically chosen points individual
threads migrate from the mobile device to a device clone in a
cloud. There the thread executes, possibly accessing native features of
the hosting platform such as the fast CPU, network, hardware accelerators,
storage, etc.  Eventually, the thread returns back to the mobile device,
along with any state it created abroad, which it merges back into the
original process. The choice of where to migrate off and back onto the
mobile device is made by a partitioning component, which uses
static analysis to discover constraints on possible migration points,
and dynamic profiling to build a cost model for execution and
migration. A mathematical optimizer chooses  migration points
that optimize execution time given the
application and the cost model.
Figure~\ref{fig:architecture} shows
the high-level architecture of our prototype.

Much research has attacked application partitioning and migration in the
past (we present detailed related work in
Section~\ref{sec:relatedwork}). We distill our novel contributions here
as follows. First, unlike traditional suspend-migrate-resume
mechanisms~\cite{isr} for application migration, the CloneCloud migrator
operates at thread granularity, an essential consideration for mobile
applications, which tend to have features that must remain at the mobile
device, such as those accessing the camera or managing the user
interface.  Second, unlike past application-layer VM
migrators~\cite{cJVM,Jessica2}, the CloneCloud migrator allows native
system operations to execute both at the mobile device and at its clones
in the cloud, harnessing not only raw CPU cloud power, but also system
facilities  or specialized hardware.
Third, unlike mostly programmer-assisted approaches to application 
partitioning, the CloneCloud partitioner automatically identifies costs 
and constraints through static and dynamic code analysis, without the 
programmer's help, annotations, or application refactoring.

In what follows, we first give some brief background on
application-layer VMs (Section~\ref{sec:background}).
We then present the design of CloneCloud's 
partitioning components (Section~\ref{sec:partitioning}) and
its distributed execution mechanism (Section~\ref{sec:migration}).
We describe our implementation
(Section~\ref{sec:impl}) and experimental evaluation
of the prototype (Section~\ref{sec:evaluation}). We survey 
related work in Section~\ref{sec:relatedwork}, discuss 
future research agenda in Section~\ref{sec:discussion}, and 
conclude in Section~\ref{sec:conc}.

\section{Background: Application VMs}
\label{sec:background}

\begin{figure}
\centering
\includegraphics{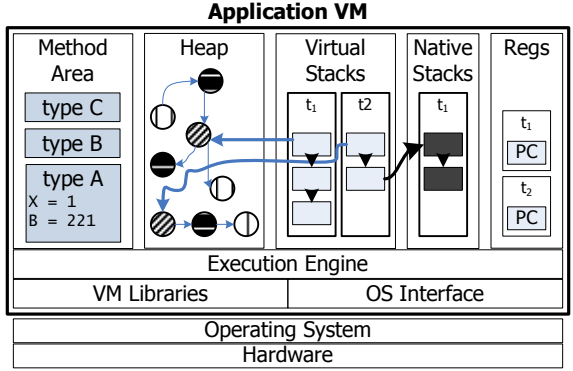}
\caption{A general architecture for an application-layer virtual
  machine.}
\label{fig:AppVM}
\end{figure}

An application-level VM is an abstract computing machine that provides
hardware and operating system independence (Figure~\ref{fig:AppVM}).
Its instruction sets are platform-independent bytecodes; an executable
is a blob of bytecodes.
The VM runtime executes bytecodes of methods with threads. There is
typically a separation between the \emph{virtual} portion of an
execution and the \emph{native} portion; the former is only expressed in
terms of objects directly visible to the bytecode, while the latter
include management machinery for the virtual machine itself, data and
computation invoked on behalf of a virtual computation, as well the
process-level data of the OS process containing the VM.
Interfacing between the virtual and the native portion happens via native
interface frameworks.

Runtime memory is split between VM-wide and per-thread areas. The
\emph{Method Area}, which contains the types of the executing program
and libraries as well as static variable contents, and the \emph{Heap},
which holds all dynamically allocated data, are VM-wide.  Each thread
has its own \emph{Virtual Stack} (stack frames of the virtual hardware),
the \emph{Virtual Registers} (e.g., the program counter), and the
\emph{Native Stack} (containing any native execution frames of a thread,
if it has invoked native functions).

Most computation, data structure manipulation, and memory management are done within the
abstract machine.  However, external processing such as file I/O,
networking, using local hardware such as sensors, are done via APIs
that punch through the abstract machine into the process's system call
interface.

\section{Partitioning}
\label{sec:partitioning}

The partitioning mechanism in CloneCloud aims to modify an application
executable by deciding where to execute methods in the code.  No special
considerations are required for the executable beyond targeting the same
application VM; that is, it need not be written in a particular idiom,
e.g., a dataflow language.  The output of the partitioning mechanism is
the executable with partitioning points, optimal for a choice of
execution conditions (network link characteristics between mobile device
and cloud, relative CPU speeds). The partitioning mechanism can be run
multiple times for different execution conditions, resulting in a
database that maps partitioning to conditions. At runtime, the
distributed execution mechanism we describe in
Section~\ref{sec:migration} implements the choice of partition for the
current execution conditions.

\begin{figure}
\centering
\includegraphics{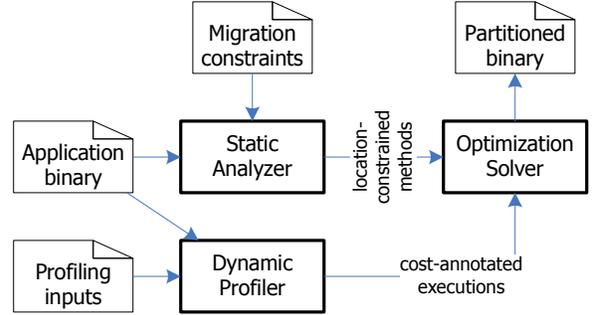} 
\caption{Partitioning analysis framework.}
\label{fig:partitioning}
\end{figure}

Partitioning of an application operates according to the conceptual
workflow of Figure~\ref{fig:partitioning}. Our partitioning framework
combines static program analysis with dynamic program profiling 
to produce partitioning that optimizes goals while meeting correctness
constraints.

The first component, the \emph{Static Analyzer}, identifies legal partition choices for
the application executable, according to a set of constraints (Section~\ref{sec:static}). 
Constraints codify the needs of the distributed execution engine used, 
as well as the particular usage model we target;
however, different mechanisms can seamlessly be plugged into the
partitioning component by changing these constraints. 

The second component, the \emph{Dynamic Profiler} (Section~\ref{sec:dynamic}),
runs the input executable on different platforms (the mobile
device and on the cloud clone) under a set of inputs, and returns a
set of profiled executions.  Profiled executions are used to
compose a cost model for the application under different
partitionings. 

Finally, the \emph{Optimization Solver} finds a legal
partitioning among those enabled by the static analyzer that minimizes
an objective function, using the cost model derived by the
profiler (Section~\ref{sec:optimizer}). 
The resulting partitioning is used to modify the executable,
yielding the final output of the partitioner. 
This partitioning is an offline process that generates a model that the runtime
uses.

\subsection{Static Analyzer}
\label{sec:static}

\begin{figure}
\centering
\includegraphics{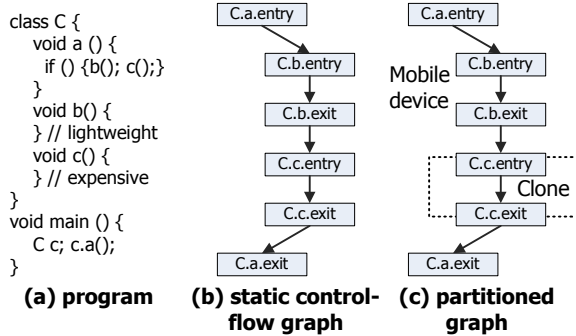} 
\caption{An example of a program, its corresponding static control-flow graph, and a partitioning}
\label{fig:pcfg}
\end{figure}

The partitioner uses static analysis to identify legal choices
for placing migration and re-integration points in the code.
In principle, these points could be placed anywhere in the
code, but we reduce the available choices to make the optimization
problem tractable.
In particular, we restrict migration and re-integration points to
the entry and exit points, respectively, of methods.
In addition, to focus on our application program, we restrict 
these partitioning points to methods of application classes as opposed to methods
of system classes (e.g., the core classes for Java) or native methods.

Figure~\ref{fig:pcfg} shows an example of a program, relevant parts of
its static control-flow graph, and a particular legal partitioning of
the program.  Class \texttt{C} has three methods.  Method \texttt{a()}
calls method \texttt{b()}, which performs lightweight processing,
followed by method \texttt{c()}, which performs expensive processing.
The static control-flow graph approximates control flow in the program
(inferring exact control flow is undecidable as program reachability is
undecidable).  The approximation is conservative in that if an execution
of the program follows a certain path then that path exists in the graph
(but the converse typically does not hold).  In the depicted static
control-flow graph, only entry and exit nodes of methods are shown,
labelled as <class name>.<method name>.<entry | exit>; other kinds of
nodes (e.g. those corresponding to instructions) are omitted since we
restrict partitioning points to method entry and exit.  A possible
partitioning as shown in Figure~\ref{fig:pcfg}c runs the body of method
\texttt{c()} on the clone, and the rest of the program on the mobile
device.

\subsubsection{Constraints}
\label{sec:constraints}

We next describe three properties required by the migration component
of any legal partitioning
and explain how we use static analysis to obtain constraints that
express these properties.

\begin{prop}
Methods that access specific features of a machine must be pinned to the machine.
\end{prop}
If a method uses a local resource such as the location service (e.g.,
GPS) or sensor inputs (e.g., microphones) in a mobile device, the method
must be executed on the mobile device. This primarily concerns native
methods, but also the main method of a program.  The analysis marks the
declaration of such methods with a special annotation $M$---for Mobile
device.  We manually identify such methods in the VM's API (e.g., VM API
methods explicitly referring to the camera); this is done once for a
given platform and is not repeated for each application. We also always
mark the main method of a program.  We refer to methods marked with $M$
as the $V_M$ method set.

\begin{prop}
Methods that share native state must be colocated at the same machine.
\end{prop}
An application may have native methods that create and access state
below the VM. Native methods may share native state. Such
methods must be collocated at the same machine as our migration component
does not migrate native state (Section \ref{sec:capture}).  To avoid a
manual-annotation burden, native state annotations are inferred
automatically by the following simple approximation, which works well in
practice: we assign a unique annotation $\mathit{Nat}_{C}$ to all native
methods declared in the same class $C$; the set $V_{\mathit{Nat}_C}$
contains all methods with that annotation.

\begin{prop}
Prevent cyclic migration.
\end{prop}
With one phone and one clone, this implies that there should be no
nested suspends and no nested resumes.  Once a program is suspended for
migration at the entry point of a method, the program should not be
suspended again without a resume, i.e., migration and re-integration
points must be executed alternately.  To enforce this property, the
static analysis builds the static control-flow graph of an application,
capturing the caller-callee method relation; it exports this as two
relations, $DC(m_1,m_2)$, read as ``method $m_1$ $D$irectly Calls method
$m_2$,'' and $TC(m_1,m_2)$ read as ``method $m_1$ $T$ransitively Calls
method $m_2$,'' which is the transitive closure of $DC$.  For the
example in Figure \ref{fig:pcfg}, this ensures that if partitioning
points are placed in \texttt{a()} then they are not placed in
\texttt{b()} or \texttt{c()}. The other remaining legal partitionings
place no migration points at \texttt{a()} but at \texttt{b()}, at
\texttt{c()}, or at both \texttt{b()} and \texttt{c()}.

\subsection{Dynamic Profiler}
\label{sec:dynamic}

The job of the profiler is to collect the data that will be used to
construct a cost model for the application under different execution
settings.  The cost metric can be different things, including energy
expenditure, resource footprint, etc.; we focus on execution time in the
prototype presented here. 

The profiler is invoked on multiple executions of the application, each
using a different set of input data (e.g., command-line arguments and
user-interface events), and each executed once on the mobile device and
once on the clone in the cloud.  The profiler outputs a set $S$ of
executions, and for each execution a \emph{profile tree} $T$ and $T'$,
from the mobile device and the clone, respectively.

A profile tree is a compact representation of an execution on a
single platform. It is a tree with one node for each method
invocation in the execution; it is rooted at the starting
(user-defined) method invocation of the application (e.g.,
\texttt{main}).  Specific method calls in the execution are represented
as edges from the node of the caller method invocation (parent) to the
nodes of the callees (children); edge order is not important. Each node
is annotated with the cost of its particular invocation in the cost
metric (execution time in our case). In addition to its called-method
children, every non-leaf node also has a leaf child called its
\emph{residual node}. The residual node $i'$ for node $i$ represents the
residual cost of invocation $i$ that is not due to the calls invoked
within $i$; in other words, node $i'$ represents the cost of running the
body of code excluding the costs of the methods called by it. Finally, each edge is annotated with the state size at the
time of invocation of the child node, plus the state size at the end of
that invocation; this would be the amount of data that the migrator
(Section~\ref{sec:capture}) would need to capture and transmit in both
directions, if the edge were to be a migration point. Edges between a
node and its residual child have no cost.

\begin{figure}
\centering
\includegraphics{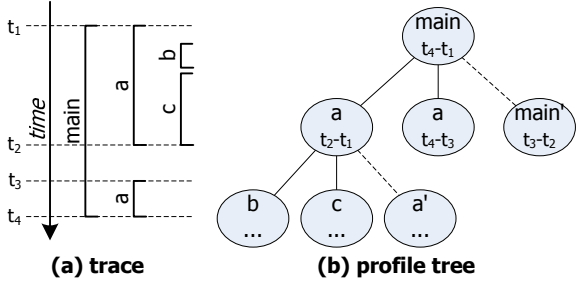} 
\caption{An example of an execution trace (a) and its corresponding
  profile tree (b). Edge costs are not shown.}
\label{fig:ProfileTree}
\end{figure}

Figure~\ref{fig:ProfileTree} is an example of an execution trace and its
corresponding profile tree.  \texttt{a} is called twice in
\texttt{main}, one \texttt{a} call invoking \texttt{b} and \texttt{c},
and one \texttt{a} call invoking no other method. A tree node on the
right holds the execution time of the corresponding method in the trace
(the length of the square bracket on the left).  \texttt{main'} and
\texttt{a'} are residual nodes, and they hold the difference between the
value of their parent node and the sum of their sibling nodes. For
example, node \texttt{main'} holds the value $t_3 - t_2 = (t_4 - t_1) -
((t_4 - t_3) + (t_2 - t_1))$.

To fill in profile trees, we temporarily instrument method entry and
exit points during each profile run on each platform. We focus only on
application code to have low profiling overhead; 
we treat system or library methods as inline code
executed in the body of the calling application method.  For our
execution-time cost metric, we collect timings at method entry and exit
points, which we process trivially to fill in tree node annotations.
For profile trees executed at the clone, we leave edge costs set to 0
 \note{rewrite - a reviewer does not understand}
(since those do not initiate migration). For mobile-device trees, we
perform the suspend-and-capture operation of the migrator 
(Section~\ref{sec:capture}), measure the
state size, and discard the captured state, both when invoking
the child node and when returning from it. Recall that for every
execution $E$, we capture two profile trees, one per platform with different
annotations.

For each invocation $i$ in profiling execution $E$, we define a computation
cost $C_c(i, l)$ and a migration cost $C_s(i)$, where $l$ is the
location of the invocation.  We fill in $C_c(i, l)$ from the corresponding
profile tree collected at location $l$; if $i$ is a leaf profile tree
node, we set $C_c(i, l)$ to be the annotation of that node; otherwise,
we set it to the annotation of the residual node $i'$.  We fill $C_s
(i)$ as the cost of making invocation $i$ a migrant invocation. 
This cost is the sum of a suspend/resume cost and a transfer cost.
The former is the time required to suspend a thread and resume a thread. 
The latter is a volume-dependent cost, the time it takes to capture,
serialize, transmit, deserialize, and reinstantiate state of a
particular size (assuming for simplicity all objects have the same such cost per byte).\note{explain how we collect this info from profiles}
We precompute this per-byte cost\footnote{One could also estimate
  this per-byte cost from memory, processor, and storage speeds, as well
  as network latency and bandwidth, but we took the simpler approach of just measuring it.}, and use
the edge annotations from the mobile-device profile tree to calculate
the cost.

\subsection{Optimization Solver}
\label{sec:optimizer}

The purpose of our optimizer is to pick which application methods to
migrate to the clone from the mobile device, so as to minimize the
expected cost of the partitioned application. Given a particular
execution $E$ and its two profile trees $T$ on the mobile device and
$T'$ on the clone, one might intuitively picture this task as optimally
replacing annotations in $T$ with those in $T'$, so as to minimize the
total node and weight cost of the hybrid profile tree.  Our static analysis
dictates the legal ways to fetch annotations from $T'$ into $T$, and our
dynamic profiling dictates the actual trees $T$ and $T'$. We do not
differentiate among different executions $E$ in the execution set $S$;
we consider them all equiprobable, although one might assign non-uniform
frequencies in practice to match a particular expected workload.

More specifically, the output of our optimizer is a value assignment to
binary decision variables $R(m)$, where $m$ is every method in the
application. If the optimizer chooses $R(m)=1$ then the partitioner will
place a migration point at the entry into the method, and a
re-integration point at the exit from the method.  If the optimizer
chooses $R(m)=0$, method $m$ is unmodified in the application
binary. For simplicity and to constrain the optimization problem, our
migration strategy chooses to migrate or not migrate \emph{all
  invocations} of a method. Despite its simplicity, this conservative
strategy provides us with undeniable benefits
(Section~\ref{sec:evaluation}); we leave further refining
differentiations depending on calling stack, method arguments, etc., to
future work.

Not all partitioning choices for $R(.)$ are
legal (Section~\ref{sec:constraints}). To express these constraints in the
optimization problem, we define an auxiliary decision variable $L(m)$
indicating the location of every method $m$, and three relations $I$, as
well as $DC$ and $TC$ computed during static analysis.  $I(i,m)$ is read
as ``$i$ is an invocation of method $m$,'' and is trivially defined from
the profile runs.  Whereas $DC$ and $TC$ are computed once for each
application, $I$ is updated with new invocations only when the set $S$
of profiling executions changes.

Using the decision variables $R(.)$, the auxiliary decision variables
$L(.)$, the method sets $V_M$ and $V_{\mathit{Nat}_C}$ for all classes $C$ defined
during static analysis, and the relations $I$, $DC$ and $TC$ from above, we
formulate the optimization constraints as follows:
\begin{align}
L(m_1) \neq L(m_2), && \forall  m_1, m_2:\mathit{DC}(m_1,m_2)=1 \nonumber \\
                   &&           \wedge R(m_2) = 1 \label{constr:move}\\
L(m) = 0,           && \forall  m \in V_M \label{constr:pinned}\\
L(m_1) = L(m_2),    && \forall  m_1, m_2, C: m_1,m_2\in V_{\mathit{Nat}_{C}}\label{constr:dependent}\\
R(m_2) = 0,         && \forall  m_1, m_2: \mathit{TC}(m_1, m_2) = 1 \nonumber \\
                   &&           \wedge R(m_1) = 1\label{constr:nested}
\end{align}
The first is a soundness constraint. Constraint~\ref{constr:move}
requires that if a method causes migration to happen, it cannot be
collocated with its callers.  The remaining three correspond to the
three properties defined in the static
analysis. Constraint~\ref{constr:pinned} requires that all methods
pinned at the mobile device run on the mobile device (Property
1). Constraint~\ref{constr:dependent} requires that methods dependent on
the native state of the same class $C$ are collocated, at either
location (Property 2). And constraint~\ref{constr:nested} requires that
all methods transitively called by a migrated method cannot be
themselves migrated (Property 3).

The cost of a (legal) partitioning $R(.)$ of execution $E$ is defined as
follows, in terms of the auxiliary variables $L(.)$, the relation $I$
and the cost variables $C_c$ and $C_s$ from the dynamic profiler:
\begin{align*}
C(E)              = & &\mathit{Comp}(E) + \mathit{Migr}(E) \\
\mathit{Comp}(E)  = & \displaystyle\sum_{i \in E,m} & [(1-L(m))I(i,m)C_c(i,0) \\
                    &                & + L(m)I(i,m)C_c(i,1)] \\
\mathit{Migr}(E)  = & \displaystyle\sum_{i \in E,m} &R(m)I(i,m)C_s(i)
\end{align*}
$\mathit{Comp}(E)$ is the computation cost of the partitioned
execution $E$ and $\mathit{Migr}(E)$ is its migration cost. For every
invocation $i \in E$, the computation cost takes its value from the
mobile-device tree annotation $C_c(i,0)$, if the method $m$ being
invoked is to run on the mobile device, or from the clone tree
annotation $C_c(i,1)$ otherwise.  The migration cost sums the individual
migration costs of only those invocations whose methods are migration
points.

Finally, the optimization objective is to choose $R()$ so as to
minimize \(\sum_{E \in S} C(E)\).  We use a standard integer linear
programming (ILP) solver to solve this optimization problem with the
above constraints.

\section{Distributed Execution}
\label{sec:migration}
The purpose of the distributed execution mechanism in CloneCloud is to 
implement a specific partitioning of an application process running inside an
application-layer virtual machine, as determined during partitioning
(Section~\ref{sec:partitioning}).

The lifecycle of a partitioned application is as follows. When the user
attempts to launch a partitioned application, current execution
conditions (availability of cloud resources and network link
characteristics between the mobile device and the cloud) are looked up
in a database of pre-computed partitions. The lookup result is a binary,
modified with particular \emph{migration} and \emph{re-integration}
points (special VM instructions in our prototype), which is then
launched in a new process. When execution of the process on the mobile
device reaches a migration point, the executing thread is suspended and
its state (including virtual state, program counter, registers, and
stack) is packaged and shipped to a synchronized clone. There, the
thread state is instantiated into a new thread with the same stack and
reachable heap objects,
and then resumed. When the migrated thread reaches a re-integration
point, it is similarly suspended and packaged as before, and then
shipped back to the mobile device. Finally, the returned packaged thread
is merged into the state of the original process.  When conditions
change, or upon explicit user input via a simple configuration
interface, a different partition and corresponding binary can be
substituted for subsequent invocations of the application.

CloneCloud migration operates at the granularity of a thread. This
allows a multi-threaded process to off-load functionality, one
thread-at-a-time.  CloneCloud enables threads, local and migrated, to
use---but not migrate---native, non-virtualized features of the platform
on which they operate: this includes the network, unvirtualized hardware
accelerators, natively implemented API functionality (such as
expensive-to-virtualize image processing routines), etc.  In contrast,
most prior work providing application-layer virtual-machine migration
keeps native features and functionality exclusively on the original
platform, only permitting the off-loading of pure, virtualized
computation.

These two unique features of CloneCloud, thread-granularity migration
and native-everywhere operation, enable new execution
models. For example, a mobile application can retain its user interface
threads running and interacting with the user, while off-loading worker
threads to the cloud if this is beneficial. This would have been impossible
with monolithic process or VM suspend-resume migration, since the user 
would have to migrate to the cloud along with the code. Similarly, a mobile
application can migrate a thread that performs heavy 3D rendering operations
to a clone with GPUs, without having to 
modify the original application source; this would have been impossible
to do seamlessly if only migration of virtualized computation were
allowed.

CloneCloud migration is effected via three distinct components: (a) a
per-process \emph{migrator thread} that assists a process with the
mechanics of suspending, packaging, resuming, and merging thread state,
(b) a per-node \emph{node manager} that handles node-to-node
communication of packaged threads, clone image synchronization and
provisioning; and (c) a simple partition database that determines what
partitioning to use.

The migrator functionality manipulates internal
state of the application-layer virtual machine; consequently we chose to
place it within the same address space as the VM, simplifying the
procedure significantly.  A manager, in contrast, makes more sense as a
per-node component shared by multiple applications, for several
reasons. First, it enables application-unspecific node maintenance,
including file-system synchronization between the device and the
cloud. Second, it amortizes the cost of communicating with the cloud
over a single, possibly authenticated and encrypted, transport
channel. Finally, it paves the way for future optimizations such as
chunk-based or similarity-enhanced data transfer~\cite{Tolia2006,
  lbfs:sosp01}.  Our current prototype has a simple configuration
interface that allows the user to manually pick out a partition from the
database, and to choose new configurations to partition for. We next
delve more deeply into the design of the distributed execution
facilities in CloneCloud.

\begin{figure}
\centering
\includegraphics{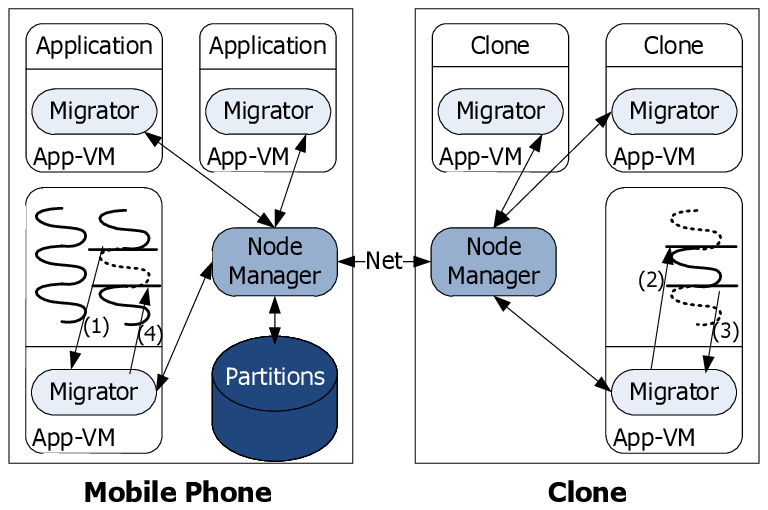} 
\caption{\label{fig:migration} Migration overview.\note{maybe cut this out}}
\end{figure}

\subsection{Suspend and Capture}
\label{sec:capture}
Upon reaching a migration point, the job of the thread migrator is to
suspend a migrant thread, collect all of its state, and pass that state
to the node manager for data transfer.  The thread migrator is a native
thread, operating within the same address space as the migrant thread,
but outside the virtual machine. As such, the migrator has the ability
to view and manipulate both native process state and virtualized state.

To capture thread state, the migrator must collect several distinct
data sets: execution stack frames \note{why multiple stack frames? - stack
walks for security policies}
and relevant data objects in
the process heap, and register contents at the migration point.
Virtualized stack frames---each containing register contents and local
object types and contents---are readily accessible, since they are
maintained by the VM management software. Starting with local data objects
in the collected stack frames, the migrator recursively follows
references to identify all relevant heap objects, in a manner similar to
any mark-and-sweep garbage collector.
For each relevant heap object, the migrator stores its field values, and
collects all relevant static fields as well (e.g., static class fields).

Captured state must be conditioned for transfer to be portable. First, object
field values are stored in network byte order to allow for
incompatibilities between different processor architectures. Second,
whereas typically a stack frame contains a local native pointer to the particular
class method it executes (which is not portable across address spaces or
processor architectures), we store instead the class name and method
name, which are portable.

\subsection{Resume and Merge}
\label{sec:resume}
As soon as the captured thread state is transferred to the target clone
device, the node manager passes that state to the migrator of a
newly allocated process. To resume that migrant thread, the migrator
must overlay the thread context over the clean process address
space. This overlaying process is essentially the inverse of the capture
process described in Section~\ref{sec:capture}. The executable text is
loaded (it can be found under the same filename in the synchronized file
system of the clone). Then all captured classes and object instances are
allocated in the virtual machine's heap, updating static and instance
field contents with those from the captured context.  As soon as the
address space contains all the data relevant to the migrant thread, the
thread itself is created, given the stack frames from the capture, the
register contents are filled to match the state of the original thread
at the migration point in the mobile device, and the thread is marked as
runnable to resume execution.

As described above, the cloned thread will eventually reach a
reintegration point in its executable, signaling that it should migrate
back to the mobile device. Reintegration is almost identical conceptually
to the original migration: the clone's migrator captures and packages
the thread state, the node manager transfers the capture back to the
mobile device, and the migrator in the original process is given the
capture for resumption.  There is, however, a subtle difference in this
reverse migration direction. Whereas in the
forward direction---from mobile device to clone---a captured thread context is used to create a new thread
from scratch, in the reverse direction---from clone to mobile device---the context must 
\emph{update} the original thread state to match the changes
effected at the clone. We call this process a \emph{state merge}.

A successful design for merging states in such a fashion depends on our
ability to map objects at the original address space to the objects they
``became'' at the cloned address space; object references themselves are
not sufficient in that respect, since in most application-layer VMs,
references are implemented as native memory addresses, which  look
different in  different processes,  across different
devices and possibly architectures, and tend to be reused over time for
different objects.

Our solution is an object mapping table, which is only used during state
capture and reinstantiation in either direction, and only stored while a
thread is executing at a clone. We instrument the VM to assign a
per-VM unique object ID to each data object created within the VM, using a
local monotonically increasing counter. For clarity, we call the ID
at the mobile device \texttt{MID} and at the clone \texttt{CID}.  Once
migration is initiated at the mobile device, a mapping
table is first created for captured objects, filling for each the
\texttt{MID} but leaving the \texttt{CID} null; this indicates that the
object has no clone counterpart yet. After instantiation at the clone,
the clone recreates all the objects with null \texttt{CID}s, assigning
valid fresh \texttt{CID}s to them, and remembers the local object
address corresponding to each mapping entry. At this point, all migrated
objects have valid mappings.

During migration in the reverse direction, objects that came from the
original thread are captured and keep their valid mapping. Newly
created objects at the clone have the locally assigned ID placed in
their \texttt{CID}, but get a null \texttt{MID}. Objects
from the original thread that may have been deleted at the clone are
ignored and no mapping is sent back for them. During the merge back at
the mobile device, we know which objects should be freshly created
(those with null \texttt{MID}s) and which objects should be overwritten
with the contents fetched back from the clone (those with non-null
\texttt{MID}s).  ``Orphaned'' objects that were migrated out but died at
the clone become disconnected from the thread object roots and are
garbage-collected subsequently. Note that the mapping table is
constructed and used only during capture and reintegration, not during
normal memory operations either at the mobile device or at the clone.

\begin{figure}
\centering
\includegraphics{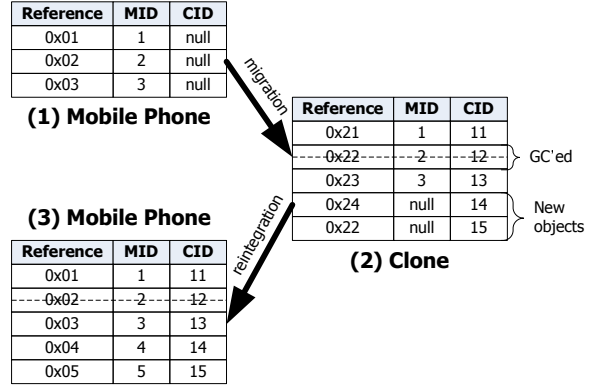} 
\caption{Object mapping example.}
\label{fig:mapping}
\end{figure}

Figure~\ref{fig:mapping} shows an example scenario demonstrating the use
of object mapping.  During initial migration, objects at addresses 0x01,
0x02, and 0x03 are captured. The migrator creates the mapping table with
three entries, one for each object, with the local ID of each
object---1, 2, and 3, respectively---in \texttt{MID}, and null
\texttt{CID}s. At the clone, the mapping table is stored, updating each
entry with the local address of each object (0x21, 0x22, and 0x23,
respectively).  When the thread is about to return back to the mobile
device, new entries are created in the table for captured objects whose
IDs are not already in the \texttt{CID} column (objects with IDs 14 and
15).  Entries in the table whose \texttt{CID} does not appear in
captured objects are deleted (the second entry in the figure). Remaining
entries belong to objects that came from the original thread and are
also going back (those with \texttt{CID} 11 and 13). Note that memory
address 0x22 was reused at the clone after the original object was
destroyed, but the object has a different ID from the original object,
allowing the migrator to differentiate between the two.  Back at the
mobile device, new objects are created for entries with null
\texttt{MID}s (bottom two entries), objects with non-null \texttt{MID}s
are updated with the returned state (first and third entries), and one
object (with local address 0x02) is left to be garbage-collected.

\subsection{Optimization}

The VM offers a unique opportunity for optimizing the amount of
information transfered during migration. Because new processes
are forked as copies of a ``template'' process ---the
\emph{Zygote}, in the Android nomenclature---and because that template
exists in all booted instances of the Android platform, we can avoid
transmitting all system heap objects that have not changed since an
 application was copied from Zygote. This typically saves about 40,000 object
transmissions with every migration operation, a significant time and
bandwidth overhead reduction. Furthermore, even ignoring the
transmission cost, some of those objects are static or
platform-dependent system objects, so should not be migrated anyway.

We obviate migration for system objects in a manner similar to how we
map objects to platform-independent IDs (in Section~\ref{sec:resume}),
with one major difference: whereas application processes are first
created at the mobile device under our control, and then partially
copied out and back in again as differences from that original single
copy, Zygote processes are created independently at the mobile device
and the clone. This creates the challenge of mapping objects from two
independent instances of Zygote on possibly different platforms. To
address the challenge, we name each Zygote object according to its
class name and invocation sequence among all objects of that
class---this assumes that objects from each class are constructed in the
same order at Zygote processes on different platforms, an assumption
that holds true in all Zygote instances we have seen so far.

\section{Implementation}
\label{sec:impl}

We implemented our prototype of CloneCloud partitioning and migration on
the ``cupcake'' branch of the Android OS~\cite{aosp}\footnote{We expect
  our design works well with other application-level VMs (e.g., JavaME)
  given the similarity of Dalvik VM and JavaME VM; testing on other
  platforms is future work.}. We tested our system on the Android Dev
Phone 1~\cite{android-dev-phone} (an unlocked HTC G1 device) equipped
with both WiFi and 3G connections, and on clones running within the
Android x86 virtual machine. We ported an ARM-based Android virtual
machine to x86 for this purpose~\footnote{Several x86-based Android VMs
  have appeared since.}. Clones execute on a Dell Desktop with a 2.83GHz
CPU and 4GB RAM, running Ubuntu 8.04.  We modified the Dalvik VM
(Android's application-level, register-based VM, principally targeted by
a Java compiler front-end) for dynamic profiling
and migration. These modifications comprised approximately 8,000 lines
of C code. We also implemented static analysis, bytecode rewriting, and
the CloneCloud node manager in Java.

For partitioning, we perform all static analysis
and bytecode rewriting with Java bytecode and convert Java bytecode into
Dalvik bytecode.  
We implemented our static analysis in jchord~\cite{jchord} and 
modified jchord to support
root methods of analysis that are different from \texttt{main}.  We
modified Dalvik VM tracing to trace migration cost and to trace only
application methods in which we are interested.  The profiling is done
both on the phone and on the clone.  Then, we use Mosek~\cite{mosek} to
solve the ILP program we defined to produce a partition for each chosen
execution environment. We use Javassist~\cite{javassist} to rewrite
bytecode to insert suspend and resume points, which are enabled or
disabled at run time depending on policies.

For migration, we modified the Dalvik VM interpreter.
For the suspend mechanism, we use Dalvik VM's implementation of thread suspension.
Each VM thread has a suspend counter which indicates if there is any pending suspend request. It checks this counter whenever it finishes the execution of a bytecode instruction, so that we can suspend the thread at a safe point. Even if a thread was executing a native frame, it also checks the counter when it finishes. The calling (migrator) thread waits until all other threads are suspended with a condition variable, and continues its execution. 

We use hprof~\cite{hprof} as a basis for capturing and representing the execution state. It provides a well-defined format for storing all the classes and heap objects efficiently. Also, since it traverses all the objects and thread stacks to collect information, we extend this format to store the thread stacks and class file paths. Also, we add the CID and MID to each object data for the mapping table. 
We implemented the object mapping table as a separate hashtable inside Dalvik VM. 
With our hashtable implementation, the hashtable is created only when migration is actually started, and destroyed after the migration. 
To track the object creation and destruction, we modified corresponding functions in Dalvik VM.

Migration is currently initiated and terminated by the (modified)
application. To pass control from the application to the migrator
thread, we define two CloneCloud APIs - \texttt{ccStart()} indicates the
start point of the migration, and \texttt{ccStop()} defines the end
point of the migration.  In partitioning, we insert these function calls
to the original application bytecode. The application thread calling
these operations notifies the migrator thread inside Dalvik, and
suspends itself. Once the migrator thread gets the notification and
gains control, it checks with the policy engine if the decision is to
migrate or not.  If the decision is yes, it handles the rest of the
migration.

\begin{table*}

\begin{small}
\begin{tabular}{||c|c|r|r|r||r|c|r||r|c|r||}
\hline
                &            & Phone     & Clone     & Max     &\multicolumn{3}{|c||}{CloneCloud 3G}& \multicolumn{3}{|c||}{CloneCloud WiFi}\\
Application     & Input      & Exec.     & Exec.     & Speedup &Exec.    & Part.   &Speedup        &Exec.           &Part.       &Speedup\\            
                & Size       &      (sec)&      (sec)& 	       &   (sec) & 	   & 		   &  (sec)  	    &            &       \\\hline\hline
Virus           &  100KB     &5.70  	 &0.30	     &	19.00  &5.70	 & Local   &1.00 	   &5.70 	    & Local	 &1.00	 \\
scanning        &  1MB       &59.70      &2.95	     &	20.24  &59.70	 & Local   &1.00 	   &20.30	    & Offload	 &2.94	 \\
                &  10MB      &640.90     &30.90	     &	20.74  & 114.52  & Offload	   & 5.60     &45.60	    & Offload	 &14.05	 \\\hline
Image           &  1 image   &22.20      &0.97	     &	22.89  &22.20	 & Local   &1.00 	   &15.90	    & Offload	 &1.40	 \\
search          &  10 images &212.20     &8.40	     &	25.26  &98.40	 & Offload &2.16 	   &23.60	    & Offload	 &8.99	 \\
                &  100 images&2096.70	 &83.20	     &	26.20  &193.10	 & Offload &10.86	   &98.90	    & Offload	 &21.20	 \\\hline
Behavior        &  depth 3   &3.60	 &0.20	     &	18.00  &3.60	 & Local   &1.00 	   &3.60 	    & Local	 &1.00	 \\
profiling       &  depth 4   &46.80	 &2.00	     &	23.40  &46.80	 & Local   &1.00 	   &14.50	    & Offload	 &3.23	 \\
                &  depth 5   &315.80	 &12.00	     &	26.32  &77.50	 & Offload &4.07 	   &25.40	    & Offload	 &12.43	 \\\hline
\end{tabular}
\end{small}

\caption{Execution times of virus scanning, image search, and behavior profiling applications. For each application we show three rows, one per input size---each application measures input size differently. For each input size, the data shown include (from left to right) execution time at the phone alone (``monolithic'' execution), execution time at the clone alone, CloneCloud execution time, partitioning choice, and speedup (for 3G connectivity), and the same information for WiFi connectivity.}
\label{tab:results}
\end{table*}

\section{Evaluation}
\label{sec:evaluation}

For the evaluation of our prototype, we implemented three mobile
applications. We evaluated running those applications either on an
Android Dev Phone 1---representing the status quo, monolithic
execution---or by optimially partitioning them for two execution
settings: one with WiFi connectivity and one with 3G connectivity.

The applications we consider are a virus scanner, image search, and
privacy-preserving targeted advertising; we briefly describe each next.
The virus scanner scans the contents of the phone file system against a
library of 1000 virus signatures, one file at a time. We vary the total size of
the file system between 100KB and 10 MB.  The image search application finds all
faces in images stored in the phone file system. We use a face-detection
Android library that returns the mid-point between the eyes, the
distance between the eyes, and the pose of every face detected. We only
use images smaller than 100KB each, due to memory limitations of the
Android face-detection library. We vary the number of images from 1 to
100. The privacy-preserving targeted advertising application uses
behavioral tracking across websites to infer the users' preferences, and
selects ads according to a resulting model; by doing this tracking at
the user's device, privacy can be protected (see
Adnostic~\cite{adnostic}). We implement Adnostic's web page
categorization on the mobile device, which maps a user's keywords to one
of the hierarchical interest categories---down to nesting levels
3-5---from the DMOZ open directory~\cite{dmoz}. The application computes
the cosine similarity between user interest keywords and predefined
category keywords.

Table~\ref{tab:results} collects all our results for the three
applications, under three different workload sizes each.  The third
column shows the execution time for each experiment when running on the
phone monolithically. As a point of comparison, the fourth column shows
execution time when the application runs on the clone in its entirety.
CloneCloud cannot achieve this performance, since in practice some part
of the application must run on the phone, and there is non-trivial
overhead in migrating the remainder to the clone. However the comparison
of these two columns, as shown in the maximum speedup column coming
next, captures the speedup opportunity available due to
the disparity between phone and cloud computation resources, when
offloading computation to a single clone.

We now turn to the choices CloneCloud makes when executing each
application using the 3G network or the WiFi network. The
execution times reported are the average of five runs.  In the 3G case,
communication is performed via an SSH tunnel between the phone and the
clone, to punch through our lab firewall.
Our 3G connection averaged latency of 415 ms, 
download bandwidth of 0.91 Mbps, and upload bandwidth of 0.16 Mbps, while our 
WiFi connection had a latency of
66 ms, download bandwidth of 7.29 Mbps, and upload bandwidth of 3.06
Mbps.\footnote{We used ping to report the average 
latency from the phone to our lab firewall, and we used Xtremelabs Speedtest,
downloaded from Android market,
to measure download and upload bandwidth.}

An obvious difference between the two execution environments is that
CloneCloud chooses to keep local more workloads (5 out of 9) in the 3G
case, than in the WiFi case (2 out of 9). This can be explained given
the overhead differences between the two networks. Migration costs about
10-15 seconds in the WiFi case, but shoots up to 60 seconds in the 3G
case, due to the greater latency \note{system would need to move
state synchronously - again state limitations clearly and argue that it
can lead a further work in this important area} and lower bandwidth in the latter
case. In both cases, migration costs include a network-unspecific
thread-merge cost---patching up references in the running address space
from the migrated thread---and the network-specific transmission of the
thread state. The former dominates the latter for WiFi, but is dominated
by the latter for 3G. A secondary effect in the results is that larger
workloads benefit from off-loading more: this is due to amortization of
the migration cost over a larger computation at the clone that receives
a significant speedup. Nevertheless, the WiFi case displays significant
speed-ups in all applications: 14x, 21x, and 12x for the largest
workload of each of the three applications, for a completely automatic
modification of the application binary without programmer input.  We
expect these benefits to increase with a number of optimizations
targeting the network overheads (in particular, 3G network overheads): 
redundant transmission elimination and compression. 

Next, we analyze the time to run the partitioning framework.
First, we report the time to perform partitioning analysis
for the image search application.
In our evaluation, we report the average of five runs.   
We profile 35 methods in the application
program. Note that we do profiling of only methods appeared in the application;
thus profiling is done with low overhead.
We profile the application on the phone and on the clone. 
Profiling execution time takes 29.4 seconds on the phone and
1.2 seconds on the clone. 
Profiling migration cost takes 98.4 seconds on the phone.
Then, running static analysis using jchord takes 19.4 seconds with 
sun jdk $1.5.0\_16$ on the desktop machine.
Generating an optimizer (ILP) script from the profile trees and constraints
and solving the generated ILP take less than one second.

\section{Related Work}
\label{sec:relatedwork}

CloneCloud is built upon previous research work done in automatic partitioning,
migration, and remote execution, and it combines these technologies 
in a non-trivial way.
First, it uses a partitioning framework that combines 
static program analysis with dynamic program profiling. It does partitioning 
in a method level, allows placing methods
that access native state remotely if they meet partitioning constraints
generated by the partitioning framework, and uses partitioning that optimizes
certain metrics. 
CloneCloud performs migrating specific threads with relevant execution state including relevant reachable heap objects. 
It performs migration on demand if doing so is beneficial, and can merge migrated state back to the original process.

\paragraph{Partitioning}
We first summarize previous work on partitioning of distributed systems.
Coign~\cite{Hunt1999} automatically partitions a distributed application composed of Microsoft COM components to reduce communication cost of partitioned components. The application must be structured to use COM
components and partitioning points are COM boundaries, and the work
focuses on static partitioning and assumes that a COM component can be
placed anywhere.
Wishbone~\cite{Newton2009} and Pleiades~\cite{Kothari2007} 
compile a central program into multiple code pieces with stubs for 
communication mostly for sensor networks.
Wishbone~\cite{Newton2009} is a system that takes an acyclic 
dataflow graph of operators written in a high-level stream-processing language and partitions the dataflow graph between server and a set of embedded nodes for sensor computing applications. It uses a compiler that generates 
partitioned source code with communication stubs based on profiling CPU and network bandwidth consumption. 
Pleiades~\cite{Kothari2007} compiles a central program written 
in an extended C language with the model of accessing the entire network
into multiple units to run on sensor nodes.
MAUI~\cite{maui} partitions applications using dynamic profiling and optimization, focusing on energy consumption. For offloaded execution, it performs method shipping with relevant heap objects.
J-Orchestra~\cite{Tilevich2002} creates partitioned applications automatically 
by a compiler that classifies anchored unmodifiable, anchored modifiable, 
or mobile classes. 
After the analysis, it
rewrites all references into indirect 
references  (i.e., references to proxy objects) for a cluster of machines,
and places classes with location constraints (e.g., ones with native state 
constraints) to proper locations.
Finally, for distributed execution of partitioned applications, it relies on the RMI middleware.

There are also Java program partitioning systems for mobile devices whose
limitation is that only Java classes without native state can be placed 
remotely~\cite{Gu2003, Ou2007, Messer2002}. The general approach is to
partition Java \textit{classes} into groups using adapted MINCUT
heuristic algorithms to minimize the component interactions between partitions. Also, different proposals consider different additional objectives  such as memory, CPU, or bandwidth. This previous work does not 
consider partitioning constraints like our work does, the granularity
of partitioning is coarse since it is a class level, and it focuses on
static partitioning.

On a related front, Links~\cite{Cooper2006}, Hops~\cite{Serrano2006}, and UML-based Hilda~\cite{Yang2006} aim to statically partition a client-server
program written in a high-level functional language or a high-level 
declarative language into two or three tiers.
Yang et. al~\cite{Yang2007} examine partitioning of programs written in Hilda based on cost functions for optimizing user response time.
Swift~\cite{Chong2007} statically partitions a program written in the Jif 
programming language into client-side and server-side computation.
Its focus is to achieve confidentiality and integrity of the partitioned program with the help of security labels in the program annotated by programmers.

\paragraph{Migration}
There has been previous work on supporting migration in Java. 
MERPATI~\cite{MERPATI} provides JVM migration using checkpointing the entire heap and all the threads with their execution environment (the call stack, the local variables, and the operand stacks) and resuming from a checkpoint.
In addition, there has been different approaches on distributed Java virtual machines (DJVMs). They assume a cluster environment where homogeneous machines are connected via fast interconnect, and try to provide a single system image to users. 
One approach is to build a DJVM upon a cluster enabled infrastructure below the JVM.
Jessica~\cite{Jessica} and Java/DSM~\cite{JavaDSM} rely on page-based distributed shared memory (DSM) systems to solve distributed memory consistency problems. To address the overhead induced by false sharing in page-based DSM systems, Jessica2~\cite{Jessica2} propose an object-based solution.  
cJVM~\cite{cJVM} implements a DJVM by modifying JVM to support method shipping to remote objects with proxy objects, creating threads remotely, and supporting distributed stacks. 
Object migration systems such as Emerald~\cite{emerald} move objects to the sites running threads requesting to access the objects. 
In contrast, CloneCloud migration chooses partial threads to offload, moves only their relevant, sufficient execution state (thread stack and relevant reachable heap objects), and supports merging between existing state and migrated execution state.

\paragraph{Remote execution}
Remote execution of resource-intensive applications for resource-poor 
hardware is a well-known approach in mobile/pervasive computing. 
All remote execution work carefully designs and pre-partitions applications between local and remote execution. Typical remote execution systems run a simple visual, audio output routine at the mobile device and computation-intensive jobs at a remote server~\cite{Fox1996,Rudenko1998,Flinn1999,Flinn2001,Young2001,Balan2002}.
Rudenko et al.~\cite{Rudenko1998} and Flinn and Satyanarayanan~\cite{Flinn1999} explore saving power via remote execution. 
Cyber foraging~\cite{Balan2002,Balan2007} uses surrogates (untrusted and unmanaged
public machines) opportunistically to improve the performance of 
mobile devices. For example, both data staging~\cite{Flinn2003} and Slingshot~\cite{Su2005} use surrogates. In particular, Slingshot creates 
a secondary replica of a home server at nearby surrogates.
ISR~\cite{isr} provides the ability to suspend on
one machine and resume on another machine by storing virtual machine (e.g., Xen) images in a distributed storage system.

Finally, our work takes a step towards achieving the vision presented in
an earlier workshop paper\cite{Chun2009}, where we made the case for
augmented smartphone execution through clones running in the cloud. In
this paper, we have presented the concrete design, implementation, and
 evaluation of our prototype system for such execution.

\section{Discussion and Future Work}
\label{sec:discussion}

CloneCloud is limited in some respects by its inability to migrate
native state and to export unique native resources remotely.
Conceptually, if one were to migrate at a point in the execution in
which a thread is executing native code, or has native heap state, the
migrator would have to collect such native context for transfer as
well. However, the complexity of capturing such information in a
portable fashion (and the complexity of integrating such captures after
migration) is significantly higher, given processor architecture differences,
differences in file descriptors, etc. As a result, CloneCloud focuses on
migrating at execution points where no native state (in the stack or the
heap) need be collected and migrated.

A related limitation is that CloneCloud does not virtualize access to
native resources that are not virtualized already and are not available
on the clone. For example, if a method accesses a camera/GPS on the mobile device, CloneCloud requires that method to remain
pinned on the mobile device.  In contrast, networking hardware or an
unvirtualized OS facility (e.g., Android's image processing API) are
available on both the mobile device and the clone, so a method that
needs to access them need not be pinned.  An alternative design would
have been to permit migration of such methods, but enable access to the
unique native resource via some RPC-like mechanism.  We consider this
alternative a complementary point in the design space, and plan to
pursue it in conjunction with thread-granularity migration in the
future.

The system presented in this paper allows only perfunctory
concurrency between the unmigrated threads and the migrated thread;
pre-existing state on the mobile device remains unmodifiable until the
migrant thread returns. As long as local threads only read existing
objects and modify only newly created objects, they can operate in
tandem with the clone. Otherwise, they have to block.  A promising
direction, whose benefits may or may not be borne out by the associated complexity, lies in extending 
this architecture to support full concurrency between the mobile device and clones.
To achieve this, we need to add thread synchronization, heap object synchronization,
on-demand object paging to access remote objects, etc. 

While in this paper we assume that the environment in which we run clone
VMs is trusted, the future of roaming devices that use clouds where they
find them demands a more careful approach.  For instance, many have
envisioned a future in which public infrastructure machines such as
public kiosks~\cite{Garriss2008} and digital signs are widely available
for running opportunistically off-loaded computations.  We plan to
extend our basic system to check that the execution done in the remote
machine is trusted.  Automatically refactoring computation around
trusted features on the clone is an interesting research question.

In our related position paper~\cite{Chun2009}, we discussed a rich
design space for automatic off-loading. Our work here covers some
aspects of primary and background augmentation, and we would like to
continue to explore hardware augmentation and multiplicity augmentation 
that uses multiple copies of the system image executed in different ways.

\section{Conclusion}
\label{sec:conc}

This paper takes a step towards seamlessly interfacing between the
mobile and the cloud in the context of mobile cloud computing.  Our
system overcomes design and implementation challenges to achieve basic
augmented execution of mobile applications on the cloud, representing
the whole-sale transfer of control from the device to the clone and
back. We combine partitioning, migration with merging, and on-demand
instantiation of partitioning to address these challenges. Our prototype
delivers up to 21.2x speedup for applications we tested, without programmer
involvement, demonstrating feasibility for the approach, and opening up
a path for a rich research agenda in hybrid mobile-cloud systems.

\bibliographystyle{abbrv}
\bibliography{clonecloud}

\end{document}